\documentstyle[amsfonts,amssymb,prl,aps,multicol,graphicx,color]{revtex}
\newcommand{\bea}{\begin{eqnarray}}
\newcommand{\eea}{\end{eqnarray}}
\newcommand{\be}{\begin{equation}}
\newcommand{\ee}{\end{equation}}

\def\llb{{\Bigg {\lbrack}}\!\!{\Bigg {\lbrack}}}
\def\rrb{{\Bigg {\rbrack}}\!\!{\Bigg {\rbrack}}}

\def\lsim{\mathrel{\lower2.5pt\vbox{\lineskip=0pt\baselineskip=0pt
          \hbox{$<$}\hbox{$\sim$}}}}
\def\gsim{\mathrel{\lower2.5pt\vbox{\lineskip=0pt\baselineskip=0pt
          \hbox{$>$}\hbox{$\sim$}}}}

\begin{document}

\begin{titlepage}
\begin{trivlist}
\item
\vspace*{4.0ex}
{\Large \textsf{MTR01W0000019}}\\[-0.8ex]
\hrule ~\\[1.8ex]
{\Large \textsf{ MITRE TECHNICAL REPORT}}\\[2.5cm]
\begin{center}
{\huge \textsf{\textbf{The Secrecy Capacity of}}}\\[0.5ex]
{\huge \textsf{\textbf{Practical Quantum Cryptography}}}\\[3.5cm]
\end{center}
{\Large \textsf{G. Gilbert}}\\[1.2ex]
{\Large \textsf{M. Hamrick}}\\[0.8ex]
~\\
{\Large \textsf{\textbf{May 2001}}}\\[3.8cm]
\begingroup\normalsize
\begin{tabbing}
{\normalsize \textsf{\textbf{Sponsor:}} \phantom{spo}} \= 
{\normalsize \textsf{The MITRE Corporation} \phantom{phantomphantomprospo}} \= 
{\normalsize \textsf{\textbf{Contract No.:}} \phantom{pro}} \= 
{\normalsize \textsf{DAAB07-01-C-C201}} \\[0.2ex]
{\normalsize \textsf{\textbf{Dept. No.:}}} \>
{\normalsize \textsf{W072}} \>
{\normalsize \textsf{\textbf{Project No.:}}} \>
{\normalsize \textsf{51MSR837}} \\[0.3cm]
\textsf{The views, opinions and/or findings contained in this report} \> \>
   \textsf{Approved for public release;} \\
\textsf{are those of The MITRE Corporation and should not be} \> \>
   \textsf{distribution unlimited.}  \\
\textsf{construed as an official Government position, policy, or} \\
\textsf{decision, unless designated by other documentation.} \\[0.3cm]
\textsf{\copyright 2001 The MITRE Corporation}
\end{tabbing}
\endgroup
~\\[0.5cm]
{\huge \textsf{\textbf{MITRE}}}\\[0.5ex]
{\large \textsf{\textbf{Washington ${\mathbf C^3}$ Center}}}\\[0.5ex]
{\large \textsf{\textbf{McLean, Virginia}}}\\
\clearpage
\end{trivlist}
\end{titlepage}

\onecolumn

\title{The Secrecy Capacity of Practical Quantum Cryptography$^{\ast}$}

\author{G. Gilbert$^{\dag}$  and M. Hamrick$^{\ddag}$}
\address{The MITRE Corporation, 12 Christopher Way, Eatontown, NJ 07724, USA }

\date{\today}

\maketitle

\begin{abstract}
Quantum cryptography has attracted much recent attention due to its potential for 
providing secret communications that cannot be decrypted by any amount of computational 
effort. This is the first analysis of the secrecy of a practical implementation of the 
BB84 protocol that simultaneously takes into account and presents the {\it full} 
set of complete analytical expressions for effects due to the presence of pulses containing 
multiple photons in the attenuated output of the laser, the finite length of individual 
blocks of key material, losses due to error correction, privacy amplification,  
continuous authentication, errors in polarization detection, the efficiency of the
detectors, and attenuation processes in the transmission medium.
The analysis addresses eavesdropping attacks on individual 
photons rather than collective attacks in general.  
Of particular importance is the first derivation of the necessary and sufficient 
amount of privacy amplification compression to ensure secrecy against the loss 
of key material  
which occurs when an eavesdropper makes optimized individual attacks on 
pulses containing multiple photons.  It is shown that only a fraction 
of the information in the multiple photon pulses is actually lost to the eavesdropper.  
\end{abstract}

\begin{multicols}{2}[]

The use of quantum cryptographic protocols to generate key material for use in 
the encryption of classically transmitted messages has been the subject of intense 
research activity.  The first such protocol, known as BB84 
\cite{bb84}, can be realized by encoding the quantum bits representing 
the raw crytpographic key  
as polarization states of individual photons.  The protocol results in the generation 
of a shorter string of key material for use by two individuals, conventionally designated 
Alice and Bob, who wish to communicate using encrypted messages which cannot be 
deciphered by a third 
party, conventionally called Eve.  The unconditional secrecy of BB84 has been proved 
under idealized conditions, namely, on the assumption of pure single-photon sources and in 
the absence of various losses introduced by the equipment which generates and detects the 
photons or by the quantum channel itself \cite{idealproof}.
The conditions under which secrecy can be maintained 
under more realistic circumstances have been studied extensively 
\cite{lutkenhaus-practical,practical-1,practical-2,practical-3}. 
This is the first analysis of the secrecy of a practical implementation of the 
BB84 protocol that simultaneously takes into account and presents the {\it full} 
set of complete analytical expressions for effects due to the presence of pulses containing 
multiple photons in the attenuated output of the laser, the finite length of individual 
blocks of key material, losses due to error correction, privacy amplification,
continuous authentication, errors in polarization detection, the efficiency of the
detectors, and attenuation processes in the transmission medium \cite{gh}.
We consider in this paper attacks made on individual
photons, as opposed to collective attacks on the full quantum state of 
the photon pulses.  
The extension to other protocols, such as 
B92 \cite{b92} is straightforward, but is not discussed here due to limitations of space.    

The protocol begins when Alice selects a random string of $m$ bits from 
which Bob and she will 
distill a shorter key of $L$ bits which they both share and 
about which 
Eve has exponentially small information.  We define the secrecy capacity ${\cal S}$ as 
the ratio of the length of the final key to the length of the original string: 

\be
\label{E:S}
{\cal S} = {L \over m}
\ee

\noindent
This quantity is useful for two reasons.  First, it can be used in proving the 
secrecy of specific practical quantum cryptographic protocols by establishing that

\be
{\cal S} > 0
\ee

\noindent
holds for the protocol.  Second, it can be used to establish the rate of generation 
of key material according to 

\be
{\cal R} = {{\cal S} \over\tau}~,
\ee

\noindent
where $\tau$ is the pulse period of the initial sequence of photon transmissions.  
Several scenarios in which useful key generation rates can be obtained are 
described in \cite{gh}.

The length of the final key is given by

\be
\label{E:L}
L = n - \left( e_T + q + t + \nu \right) - \left( a + g_{pa} \right)
\ee

\noindent
The first term, $n$ is the length of the sifted string.  This is the string that 
remains after Alice has sent her 
qubits to Bob, and Bob has informed Alice of which qubits were received and in what 
measurement 
basis, and Alice has indicated to Bob which basis choices correspond to her own.  
We consider here the important special case where the number of photons in the pulses 
sent by Alice 
follow a Poisson distribution with parameter $\mu$.  
This is an appropriate description when the source is a pulsed laser 
that has been attenuated to produce weak coherent pulses.  
In this case, the length of the sifted string may be expressed as \cite{gh}

\be
\label{E:n}
n = 
{m\over 2}{\Bigg [}\psi_{\ge 1}\left(\eta\mu\alpha\right) \left( 1 - r_d \right)
+r_d{\Bigg ]}~,
\ee

\noindent
where $\eta$ is the efficiency of Bob's detector, $\alpha$ is the transmission 
probability in the quantum channel, and $r_d$ is the probability of obtaining a 
dark count in Bob's detector during a single pulse period.   
$\psi_{\ge k}\left(X\right)$ is 
the probability of encountering $k$ or more photons in a pulse selected at random from
a stream of Poisson pulses having a mean of X photons per pulse:

\be
\label{E:psi}
\psi_{\ge k}\left(X\right) \equiv
\sum_{l=k}^\infty \psi_l\left(X\right) =
\sum_{l=k}^\infty e^{-X}{X^l\over l!},
\ee

\noindent
Other types of photon sources may be treated by appropriate modifications 
of equations \ref{E:n} and \ref{E:psi}.  
A comprehensive treatment of 
this subject, including an extensive analysis of factors contributing to $\alpha$, is found 
in \cite{gh}.  

The next terms represent information that is either in error, or that may potentially 
be leaked to Eve during the rest of the protocol.  This information is removed 
from the sifted string by the algorithm used for privacy amplification, 
and so the corresponding number of bits must be subtracted 
from the length of the sifted string to obtain the size of the final key that results.  

The first such term, $e_T$, represents the errors in the sifted string.  This may be 
expressed in terms of the parameters already defined and the intrinsic channel error 
probability $r_c$:

\be
e_T = 
{m\over 2}{\Bigg [}\psi_{\ge 1}\left(\eta\mu\alpha\right) r_c \left( 1 - r_d \right)
+{r_d \over 2}{\Bigg ]}~,
\ee

\noindent
where the intrinsic channel errors are due to relative misalignment of Alice's and 
Bob's polarization axes and, in the case of fiber optics, the dispersion 
characteristics of the transmission medium.   
These errors are 
removed by an error correction protocol which results in an additional $q$ bits of 
information about the key being transmitted over the classical channel.  We express this 
as 

\bea
q&\equiv& Q\left(x,{e_T\over n}\right) e_T
\nonumber\\
&=&{xh\left(e_T/n\right)\over e_T/n} e_T
\eea

\noindent
where $h(p)$ is the binary entropy function for a bit whose {\it a priori } 
probability of being $1$ is $p$.  The factor $x$ is introduced as a measure of the ratio 
by which a particular error correction protocol exceeds the theoretical minimum amount of 
leakage given by Shannon entropy \cite{shannon}:  

\be
q_{min} = n h\left(e_T/n\right) = {h\left(e_T/n\right)\over e_T/n} e_T
\ee

The next term, $t$, is an upper bound for the amount of information Eve can obtain by 
direct measurement of the polarizations of single photon pulses.  
This upper bound may be expressed as 

\be
\label{E:t}
t = T e_T
\ee

\noindent
where $T$ is given by \cite{gh,slutsky,lutkenhaus-T}

\bea
\label{E:T}
T\left(n_1,e_T,e_{T,1},\epsilon\right)&=&
\left({n_1\over e_{T}}-{e_{T,1}\over e_T}\right)
{\bar I}_{max}^R\left({e_{T,1}\over n_1}
+\xi\left(n_1,\epsilon\right)\right)
\nonumber\\
&&+\xi\left(n_1,\epsilon\right){n_1\over e_{T}}\left(1-{e_{T,1}\over n_1}\right)^{1/2}~,
\nonumber\\
\eea

\noindent
with

\be
\label{48}
{\bar I}_{max}^R\left(\zeta\right)\equiv 1+\log_2\left[1-{1\over 2}\left({1-3\zeta\over 1-
\zeta}\right)^2\right]~,
\ee

\noindent
and $\xi$ is defined by

\be
\label{E:xi}
\xi\left(n_1,\epsilon\right)\equiv{1\over{\sqrt {2n_1}}}{\rm {erf}}^{-1}
\left(1-\epsilon\right)~.
\ee

In the above equation $\epsilon$ is a security parameter that gives
the likelihood for a successful eavesdropping attack against a single-photon pulse in
the stream.  

Finally, we have used

\be
n_1={m\over 2}{\Big [}\psi_{1}\left(\eta\mu\alpha\right) \left( 1 - r_d \right)
   +r_d{\Big ]}~,
\ee

\noindent
and

\be
e_{T,1}={m\over 2}{\Big [}r_c\psi_{1}\left(\eta\mu\alpha\right) \left( 1 - r_d \right)
   +{r_d\over 2}{\Big ]}~,
\ee

\noindent
which are the contributions to $n$ and $e_T$ from the subset of Alice's pulses for
which exactly one photon reaches Bob.  

The next term, $\nu$, is the information leaked to Eve by making attacks on pulses 
containing more than one photon.  There are a variety of possible attacks, including 
coherent attacks that operate collectively on all the photons in the pulse.  We restrict 
our attention to disjoint attacks that single out each individual photon.  Even with this 
restriction, there are a large number of alternatives.  Eve can perform a {\it direct} 
attack by making direct measurements 
of the polarization of some subset of the photons and allowing the rest to continue 
undisturbed.  
She can also perform an {\it indirect} attack by storing some of the photons until 
she learns Alice and Bob's basis choices by eavesdropping on their classical channel.  
She then measures the stored photons in the correct basis to unambiguously determine the 
value of the bit.  Finally, she can make a {\it combined} attack by using the 
two strategies in some combination.  In 
\cite{gh} it is shown that the optimum attack is always either a direct or an indirect 
attack, depending on the value of a parameter $y$, which depends on channel and detector 
characteristics and the technological capabilities attributed to Eve \cite{gh}.  
For the case of 
a fiber optic channel, it is possible in principle for Eve to replace the cable 
with a 
lossless medium, so that those pulses whose polarizations she can measure are guaranteed to 
reach Bob.  In this case we take $y=\eta$.  For the free space case, such an 
attack may not be feasible, but she can achieve a similar effect by using entanglement.  
In this version of 
the indirect attack, Eve and an accomplice located near Bob prepare pairs
of entangled photons in advance.   Eve then entangles one of these pairs 
with a photon emitted by Alice.  Her accomplice can now make measurements on the entangled 
state, gaining information about the photons at Eve's location without losing photons 
to the attenuation in the channel.  
If we allow for such 
attacks, we still have $y=\eta$.   If we do not attribute this level of technology to 
Eve, it is appropriate to take $y=\eta\alpha$.  Note also that Eve can perform direct 
attacks 
using classical optical equipment, but that the indirect attacks require 
the type of 
apparatus envisaged for quantum computers.  

There are three regions of interest.  
If $y~>~1-{1\over\sqrt 2}~~{\Big (}i.e.,~y~
\gsim~0.293{\Big )}$, the indirect attack is stronger, 
and the maximum information 
that Eve can obtain is 

\bea
\nu^{max}={m\over 2}\llb && \psi_{\ge 2}\left(\mu\right)
-\left(1-y\right)^{-1}
\nonumber\\
&&\quad\cdot\;{\Bigg \{}e^{-y\mu}-e^{-\mu}
{\Bigg [}1+\mu\left(1-y\right){\Bigg ]}{\Bigg \}}\rrb~.
\nonumber\\
\eea

\noindent
If $y~<~1-{1\over\sqrt[3] 2}~~{\Big (}i.e.,~y~
\lsim~0.206{\Big )}$, the direct attack is stronger, and Eve's information is 

\bea
\nu^{max}={m\over 2}{\Bigg [}&&\psi_2\left(\mu\right)y+1
\nonumber\\
&&-e^{-\mu}{\Bigg (}\sqrt 2\sinh{\mu\over\sqrt 2}+2\cosh{\mu\over\sqrt 2}-1{\Bigg )}
{\Bigg ]}~.
\nonumber\\
\eea

\noindent
Finally, if $y$ lies between these two regions, the relative strength 
of the attacks depends on the number of photons in the pulse.  The information leaked to 
Eve is

\bea
\nu^{max}={m\over 2}\llb&&\psi_2\left(\mu\right)y+
e^{-\mu}{\Bigg (}\sinh\mu-{\sqrt 2}\sinh{\mu\over\sqrt 2}{\Bigg )}
\nonumber\\
&&\!\!\!\!\!\!\!\!\!\!\!\!
+\sum_{k=2}^\infty\psi_{2k}\left(\mu\right){\Bigg \{}\theta
{\Big (}\sigma_e\left(k,y\right)-1{\Big )}{\Big [}1-\left(1-y\right)^{2k-1}{\Big ]}
\nonumber\\
&&\!\!\!\!
+{\Big [}1-\theta{\Big (}\sigma_e\left(k,y\right)-1{\Big )}{\Big ]}
\left(1-2^{1-k}\right){\Bigg \}}\rrb~,
\nonumber\\
\eea

\noindent
where we have introduced the function:

\be
\sigma_e\left(k,y\right)={1-\left(1-y\right)^{2k-1}\over 1-2^{1-k}}~,
\ee

\noindent
For a photon pulse with $2k$ photons, $\sigma_e\left(k,y\right)$ is greater 
than $1$ if the 
indirect attack is stronger and less than $1$ if the direct attack is stronger.   
For odd numbers of photons, the direct attack is always stronger in this region  
\cite{gh}.  

The significance of these results for Eve is evident.  If the key distribution 
system is operating in the region of large $y$, her optimal attack is 
always the indirect attack.  
If the system operates in the region of small $y$, the direct attack is 
optimal.  If the system operates in the middle region, Eve optimizes 
her attack by measuring nondestructively the number of photons in the incoming pulses and 
then selecting the attack for each pulse according to the number of photons it contains.  

The expressions for $\nu$ represent upper bounds on the information that is leaked to Eve 
by attacks on the individual photons of multi-photon pulses.  In \cite{gh} it is 
shown that Eve can always choose an eavesdropping strategy to achieve this 
upper bound as long as Bob does not counterattack by monitoring the 
statistics of multiple detection 
events that occur at his device.  Even with this proviso, the upper bounds are only a 
fraction of the information contained in the multi-photon pulses.  
This indicates that the assumption, common in the literature, that Alice and Bob 
must surrender 
all of this information to Eve is overly restrictive.  

The next two terms are grouped together at the end of the expression because their 
effect on ${\cal S}$ vanishes in the limit of large $m$.  
The first of these, $a$, is the continuous authentication cost.
This is the number of secret 
bits that are sacrificed as part of the authentication protocol to ensure that the 
classical transmissions for sifting and error correction do occur between Alice and Bob 
without any ``man-in-the-middle" spoofing by Eve.  For the authentication protocols 
described in \cite{gh}, 
the authentication cost is

\bea
\label{E:a}
a\left(n,m\right) &=&
4{\Big \{} g_{auth} + \log_2\log_2{\Big [} 2n\left(
1+\log_2m\right){\Big ]}{\Big \}}
   \nonumber\\
   &&   \qquad \cdot \log_2{\Big [} 2n\left( 1+\log_2m\right){\Big ]}
   \nonumber\\
   &&+ 4{\Big [} g_{auth}+\log_2\log_2\left( 2n\right){\Big ]} \log_2\left(
2n\right)
   \nonumber\\
   &&+ 4\left( g_{EC} + \log_2\log_2 n\right)\log_2 n
   \nonumber\\
   &&+ 4\left( g_{auth} + \log_2\log_2 g_{EC}\right)\log_2 g_{EC}
   \nonumber\\
   &&+ \tilde g_{EC}
   \nonumber\\
   &&+ 4\left( g_{auth} + \log_2\log_2 \tilde g_{EC}\right)
      \log_2 \tilde g_{EC}~.  
\eea

\noindent
The security parameters $g_{auth}, g_{EC},$ and $\tilde g_{EC}$ are adjusted to limit 
the probability that some phase of the 
authentication fails to produce the desired result.  
For instance the probability that Eve can successfully replace Alice's transmissions 
to Bob with her own transmissions is bounded by $2^{-g_{auth}}$.  The probability that 
Alice's and Bob's copies of the key do not match after completion of the protocol 
is bounded by $2^{-g_{EC}} + 
2^{-\tilde g_{EC}}$.  

The last term, $g_{pa}$, is a security parameter that characterizes the effectiveness of 
privacy amplification.  It is the number of bits that must be sacrificed to limit the 
average amount of information, $\langle I \rangle$, about Alice and Bob's 
shared key that Eve 
can obtain to an exponentially small number of bits \cite{gpa}:

\be
\label{E:I}
\langle I \rangle \leq {{2^{-g_{pa}}}\over{\ln 2}}~.
\ee

The fundamental expression for the secrecy capacity may now be written in the limit of 
small dark count, $r_d << 1$:

\bea
{\cal S} &=& 
{1\over 2}{\Bigg [}\psi_{\ge 1}\left( \eta\mu\alpha \right) \cdot
\left(1-fr_c\right)+\left(1-{f\over 2}\right)r_d
-\tilde\nu{\Bigg ]}  \nonumber\\
&&-{g_{pa}+a\over m}~,
\eea

\noindent
where we have defined 

\be
f\equiv 1+Q+T~,
\ee

\noindent
and

\be
\tilde\nu\equiv 2\nu^{max} /m~,
\ee

\noindent
so that the rescaled quantity $\tilde\nu$ is independent of $m$.  

Note that the 
pulse intensity parameter $\mu$ can be chosen to maximize the secrecy capacity 
${\cal S}$ and thus also the key generation rate ${\cal R}$.  
A detailed investigation of the optimum pulse intensity under various conditions 
of practical interest and the resulting secrecy capacities and rates 
may be found in \cite{gh}.  

We have presented results for the secrecy capacity of a practical quantum 
key distribution scheme using attenuated laser pulses to carry the quantum 
information and encoding the raw key material using photon polarizations according 
to the BB84 protocol.  This is the first analysis of the secrecy of a practical 
implementation of the BB84 protocol that simultaneously takes into account 
and presents the {\it full} set of complete analytical expressions for effects due to the
presence of 
pulses containing multiple photons in the attenuated output of the laser, 
the finite length of 
individual blocks of key material, losses due to 
error correction, privacy amplification, continuous authentication,   
errors in polarization detection, 
the efficiency of the 
detectors, and attenuation processes in the transmission medium 
for the implemantation of BB84 described in \cite{gh}.  
The transmission medium may be either free space 
or fiber optic cable.  
The results apply when eavesdropping is restricted to attacks on individual 
photons.  
The extension of these results to include 
collective attacks on multiple photon states in full generality 
is the subject of continuing research.  
Of particular importance are the 
findings that only a 
portion of the information in the multi-photon pulses need be lost to Eve and the 
identification of those regions of operation for which Eve's attack is optimized 
by choosing direct attacks, indirect attacks, or selecting the attack in real time 
based on the number 
of photons in the pulse.  The assumption, common in the literature, that Alice and Bob 
must surrender 
all of this information to Eve is overly conservative.  
A companion paper, \cite{singlephoton}, compares quantitatively the results described
here for attenuated 
laser sources with what it is achievable 
using ideal single-photon sources.  \\

\noindent {\bf
APPENDIX: NOTE ON THE SECRECY CAPACITY FOR KEYS OF FINITE LENGTH}  \\

Most of the terms appearing in eq.(\ref{E:L}) for the length of the secret key, $L$, are 
directly proportional to the length of the block of raw key material, $m$.  After dividing 
through by $m$ ({\it cf} eq.(\ref{E:S})), the contributions of these 
terms to the secrecy capacity ${\cal S}$ are independent of $m$.  Three of the 
terms in $L$ are not proportional to $m$, namely $g_{pa},~a,~{\rm and}~t$.  
They result in contributions 
to the effective secrecy capacity that retain explicit dependence on $m$.  

The third contribution, $t$, requires additional explanation.  Its $m$ dependence 
arises from a precise application of the 
privacy amplification result, eq.(\ref{E:I}), derived by Bennett {\it et al.\/} in 
\cite{gpa}.  
The bound on Eve's knowledge 
of the final key is obtained by assuming she has obtained a specific amount of 
Renyi information prior to privacy amplification.  Starting from this point, 
Slutsky {\it et al.\/} 
\cite{slutsky} explicitly introduce a security parameter $\epsilon$ 
(see eq.(\ref{E:xi})) to bound
the probability that Eve has obtained more than $t$ bits of Renyi information as a 
result of her attacks on single photon pulses.  

By contrast, the analysis of \cite{lutkenhaus-practical} introduces no 
parameter analogous to $\epsilon$.  Furthermore, the expression for the 
amount of privacy amplification compression given in \cite{lutkenhaus-practical} is 
{\it linear} in the blocksize, 
thus resulting in a contribution to the secrecy capacity that is 
independent of the blocksize.  
While this approach, as developed in \cite{lutkenhaus-practical}, 
does yield a bound on Eve's information about 
the key shared by Alice and Bob {\it after} privacy amplification, 
explicit results
pertaining to the amount of information Eve obtains on the key {\it prior} to privacy 
amplification are not presented.  Such results have important practical consequences.  
For example, Eve's likelihood of obtaining more than a given fraction 
of the raw key from her attacks on single photons increases as the block size 
of the key material is reduced.  One therefore expects that the amount of 
privacy amplification compression required to ensure secrecy will increase as well.  
However, since this conclusion is strictly a consequence of the information Eve obtains 
{\it prior} to privacy amplification, it cannot be inferred from the analysis 
of \cite{lutkenhaus-practical}.  
In contrast, the approach of \cite{slutsky},  
which we adopt in our analysis, relates the privacy amplification compression 
directly to the amount of information leaked to Eve {\it prior} to privacy amplification.   
This makes 
it possible to analyze the effect of the block size on the amount of privacy amplification 
compression, and concomitantly introduces an explicit security parameter, $\epsilon$, 
as a bound 
on Eve's chances of mounting a successful attack on strings of finite length.

\bibliographystyle{prsty}

\begin{thebibliography}{1}

\bibitem{bb84}
C. H. Bennett and G. Brassard, in {\it Proc. IEEE Int. Conference on Computers,
Systems and Signal Processing}, IEEE Press, New York (1984).

\bibitem{idealproof}
H.-K. Lo,
``A simple proof of the unconditional security of quantum key distribution,"
{\it arXive e-print} quant-ph/9904091 (1999).

\bibitem{lutkenhaus-practical}  
N. L\"utkenhaus, ``Estimates for practical quantum cryptography,"
Phys. Rev. {\bf A59}, 3301-3319 (1999).

\bibitem{practical-1}  
G. Brassard, N. L\"utkenhaus, T. Mor and B.C. Sanders,
``Security Aspects of Practical Quantum Cryptography,"
{\it arXive e-print}  quant-ph/9911054 (1999).

\bibitem{practical-2}  
B. Slutsky, P.-C. Sun, Y. Mazurenko, R. Rao and Y. Fainman,
``Effect of channel imperfection on the secrecy capacity of a quantum
cryptographic system,"
J. Mod. Opt. {\bf 44}, 5, 953-961 (1997).  

\bibitem{practical-3}  
S. F\'elix, N. Gisin, A. Stefanov, and H. Zbinden,  
``Faint laser quantum key distribution: Eavesdropping exploiting multiphoton pulses,"
{\it arXive e-print}  quant-ph/0102062 (2001).  

\bibitem{gh} 
G. Gilbert and M. Hamrick, ``Practical Quantum Cryptography: A Comprehensive 
Analysis (Part One)," {\it arXive e-print}  quant-ph/0009027 (2000).  

\bibitem{b92} 
C. H. Bennett, ``Quantum cryptography using any two nonorthogonal states," 
Phys. Rev. Lett. {\bf 68}, 3121 (1992).  

\bibitem{shannon}
C.E. Shannon,
``Communication Theory of Secrecy Systems,"
Bell Syst. Tech. J. {\bf 28}, 656 (1949).
  
\bibitem{slutsky}
B. Slutsky, R. Rao, P.-C. Sun, L. Tancevski and S. Fainman,
``Defense frontier analysis of quantum cryptographic systems,"
App. Opt. {\bf 37}, 14, 2869-2878 (1998).  

\bibitem{lutkenhaus-T}
N. L\"utkenhaus, ``Security against individual attacks for realistic quantum key 
distribution," {\it arXive e-print}  quant-ph/9910093 (2000).  

\bibitem{gpa}
C. H. Bennett, G. Brassard, C. Cr\'epeau, and U. Maurer, 
``Generalized Privacy Amplification," IEEE Trans. Inf. Th. {\bf 41}, 1915 (1995).

\bibitem{singlephoton} 
G. Gilbert and M. Hamrick, ``Single photon sources and pulsed laser sources in 
quantum cryptography," to appear.  

\end{thebibliography}

\vspace*{.175in}
{\footnotesize
$\!\!\!\!\!\!\!\!\!\!\ast$ This research was supported by MITRE under MITRE Sponsored
Research Grant
51MSR837.\\
\dag ~ggilbert@mitre.org\\
\ddag ~mhamrick@mitre.org
}

\end{multicols}
 
\end{document}